\documentclass[12pt,preprint]{aastex}

\begin{document}

\title{Evidence of Explosive Evaporation in a Microflare Observed by Hinode/EIS}

\author{F. Chen and M. D. Ding}
\affil{Department of Astronomy, Nanjing University, Nanjing 210093, China;}
\affil{Key Laboratory for Modern Astronomy and Astrophysics (Nanjing
University), Ministry of Education, Nanjing 210093, China}
\email{dmd@nju.edu.cn}

\begin{abstract}
We present a detailed study of explosive chromospheric evaporation during a microflare which occurred on 2007 December 7 as observed with the EUV Imaging Spectrometer (EIS) onboard {\it Hinode}. We find temperature-dependent upflows for lines formed from 1.0 to 2.5 MK and downflows for lines formed from 0.05 to 0.63 MK in the impulsive phase of the flare. Both the line intensity and the nonthermal line width appear enhanced in most of the lines and are temporally correlated with the time when significant evaporation was observed. Our results are consistent with the numerical simulations of flare models, which take into account a strong nonthermal electron beam  in producing the explosive chromospheric evaporation. The explosive evaporation observed in this microflare implies that the same dynamic processes may exist in events with very different magnitudes.
\end{abstract}


\section{Introduction}\label{intro}
The well-known CSHKP model \citep{car64,stu68,hira74,kopp76} for two-ribbon flares states that the flares result from magnetic reconnection in the corona, accompanied by acceleration of energetic electrons and the bulk heating of coronal plasma. The nonthermal electrons produced in this process travel along the magnetic field lines toward the cold and dense chromosphere, in which they produce hard X-ray (HXR) emission by bremsstrahlung and heat the chromospheric plasma by Coulomb collisions. In addition, direct thermal conduction from the corona can also contribute to the heating of the chromosphere. The heated chromospheric material then expands upward to the coronal loop and finally forms the hot soft X-ray (SXR) flare loop.

Such a process, which was first described by \citet{neu68}, is now known as chromospheric evaporation. The blue shift in extreme ultraviolet (EUV) and SXR emission lines, evidence of this dynamic phenomenon, was observed by \citet{act82}, \citet{anto82}, and \citet{anto83} using Bent and Bragg Crystal Spectrometer (BCS) on board the {\it Solar Maximum Mission} ({\it SMM}; \citealt{act80}), by \citet{maris93}, \citet{maris94}, \citet{ding96}, and \citet{dosc05} using BCS on board  {\it Yohkoh} \citep{cul91}, by \citet{cza01}, \citet{bro04}, and \citet{mil06a,mil06b} using Coronal Diagnostic Spectrometer (CDS; \citealt{harrison95}) on board the {\it Solar and Heliospheric Observatory} ({\it SOHO}), and recently by \citet{mil09} using the EUV Imaging Spectrometer (EIS; \citealt{cul07}) on {\it Hionde}. It is accepted that the evaporation is considered gentle when the emission lines formed in the upper chromosphere, the transition region, and the hot corona all appear to be blue shifted. The process is said to be explosive when the lines formed in the upper chromosphere and transition region appear to be red shifted, while lines with higher formation temperatures appear to be blue shifted \citep{fish85b,bro03,bro04,mil06a,mil06b,bro09,mil09,raf09}. The redshifts are thought to correspond to  chromospheric condensation caused by the pressure gradient induced by the violent heating of the chromosphere or the transition region by the energetic electrons. Conversion from an explosive to a gentle chromospheric evaporation during flares was reported by \citet{bro09}. \citet{fish85a} performed a numerical simulation on the radiative hydrodynamic process of the flare atmosphere by thick-target electron heating. They found that an electron energy flux of $10^{10}$~ergs~cm$^{-2}$~s$^{-1}$ acts as a threshold, above which the chromospheric evaporation becomes explosive. When the energy flux falls below this threshold, the evaporation is gentle, in which the chromospheric plasma is depleted at a slower rate. This theory was verified by \citet{mil06a, mil06b}, who found within the CDS and the {\it Reuven Ramaty High-Energy Solar Spectroscopic Image} ({\it RHESSI};  \citealt{lin02}) observations both a gentle chromospheric evaporation, in which the energy flux of the electron beam was below the threshold, as well as an explosive one, in which the energy flux was above the threshold.

Most of the observations mentioned above were focused on energetic events, since the velocity amplitude of the evaporated mass flows is expected to be strongly related to the energy flux of nonthermal electrons. There have been very few studies of chromospheric evaporation in small flares or microflares. \citet{mil08} reported weak blueshifts in the \ion{Fe}{15} line and \citet{broh09} found a clear case of gentle evaporation in microflares. In addition, \citet{broh07} reported evidence of explosive evaporation in a small flare-like transient, although it may have been related to a nearby M-class flare. \citet{sto07,sto08}, using {\it RHESSI} data, found that the spectra of microflares can be fitted with a thermal component at low energies and a nonthermal power law component at higher energies. Furthermore, they found that the intensity enhancement at 1600 \AA~and 171 \AA~wavebands, observed by the {\it Transition Region And Coronal Explorer} ({\it TRACE}; \citealt{han99}), is simultaneous with the peak of the HXR light curve. \citet{qiu04} analyzed HXR and microwave emissions from microflares and corroborated the nonthermal character of these emissions. Therefore, these results suggest that the scenario of chromospheric heating predicted by the standard flare model should also apply to microflares.

In this paper, we present evidence of explosive chromospheric evaporation in a microflare with {\it GOES} Soft X-ray class of B1.4. This implies that the electron flux in microflares is not always low but may sometimes be comparable to that in much more energetic events. This also suggests that the same physical process exists in active events with quite different magnitudes. We give a description of the observations of this event in \S \ref{obser} and present our data processing method in \S \ref{data}. The results are shown in \S \ref{result}. We discuss the implication of our results and present a summary in \S \ref{discu}.

\section{Observations}\label{obser}
The B1.4 two ribbon flare occurred at $\sim$~4:30 UT on 2007 December 7 in NOAA AR 10977 near the solar disk center. The flare was observed by the Extreme-ultraviolet Imaging Telescope (EIT; \citealt{dela95}) aboard the {\it SOHO} spacecraft and the Solar Optical Telescope (SOT; \citealt{tsu08}) aboard the {\it Hinode} spacecraft. The active region was scanned by EIS on {\it Hinode} repeatedly for the entire duration of the flare. The EIS observation of this event started at $\sim$~04:30 UT, using a 1\arcsec~slit that rastered with a step of 1\arcsec~and an exposure time of 30 s. The time gap between successive exposures is 2 s. The field of view (FOV) is 5\arcsec~in the scanning direction and 240\arcsec~in the slit direction. Therefore, a high raster cadence of 160 s for EIS was achieved, by sacrificing the spatial FOV. Scanning over the active region was repeated 40 times, covering the whole evolution of the flare. In addition, the same active region was rastered with EIS from 03:27 to 04:19 UT, using the 1\arcsec~slit with a step of 3\arcsec~and an exposure time of 50 s. This observation corresponds to the pre-flare phase. The FOV is 180\arcsec~in the scanning direction and 512\arcsec~in the slit direction.

\begin{figure}
\epsscale{1}
\plotone{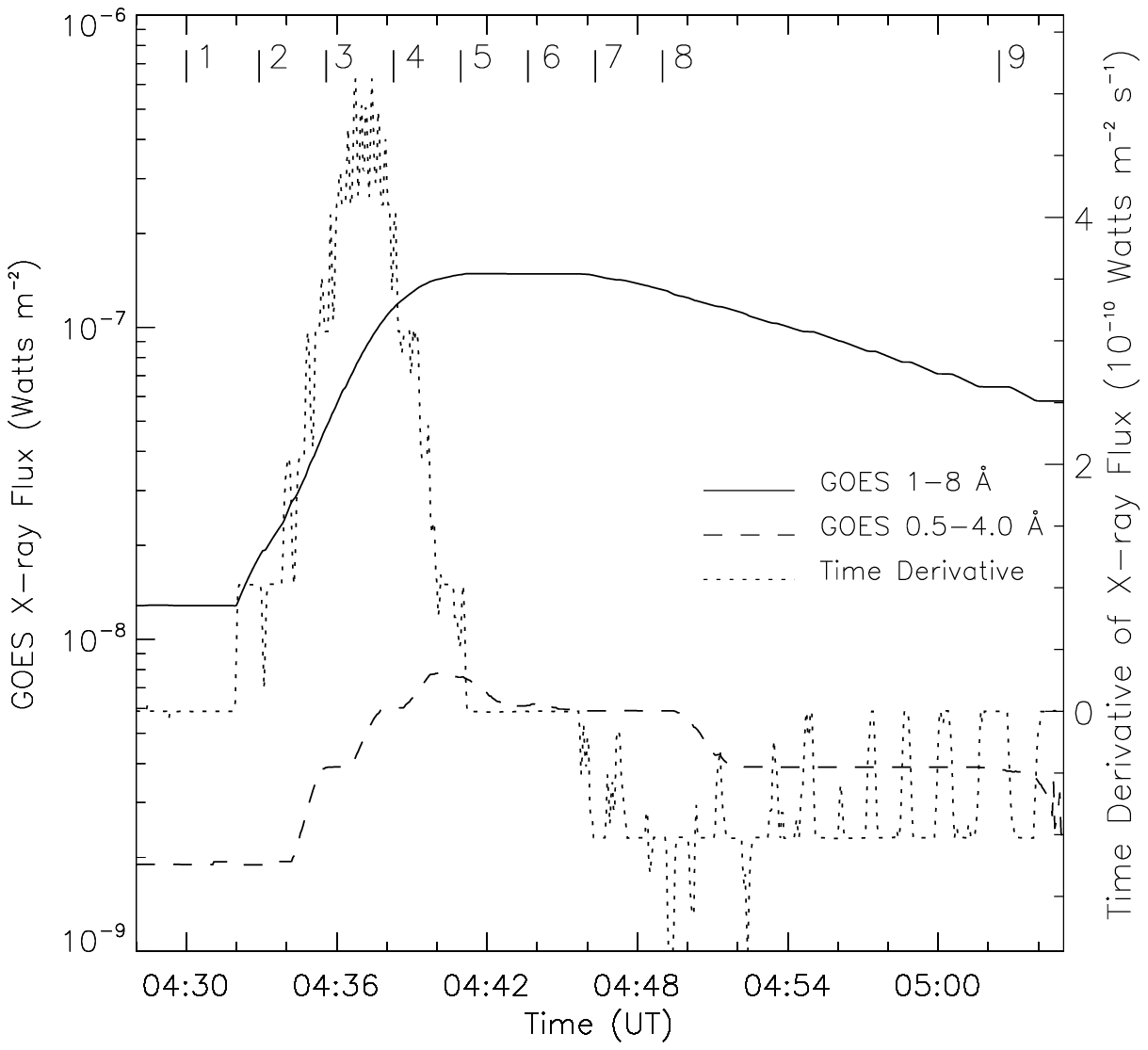}
\caption{{\it GOES} light curves for the 1--8 \AA~waveband (solid line) and the 0.5--4.0 \AA~waveband (dashed line) and the time derivative (dotted line) for the 1--8 \AA~curve. The impulsive phase of the flare is supposed to correspond to the peak of the differential curve, as implied by the Neupert effect. The vertical bars labeled with numbers refer to timings of EIS rasters.}\label{fig1}
\end{figure}

Unfortunately, there is no observation by {\it RHESSI} for this event. Instead, we use the derivative {\it GOES} 1--8 \AA~light curve as a proxy of the HXR light curve, assuming that the Neupert effect  \citep{neu68,kah70,hud91} is applicable. The {\it GOES} 1--8 \AA~ and 0.5--4 \AA~ light curves and the derivative of the former are plotted in Figure \ref{fig1} to estimate the impulsive phase of the flare. From this figure, the flare onset is at $\sim$~04:30 UT, the impulsive phase is at $\sim$~04:35--04:40 UT, and the flare peak is at $\sim$~04:40--04:45 UT, followed by the decay phase.

To compare data sets from different wavebands, we first co-align images from different instruments. The EIT 195 \AA~image is co-aligned with the pre-flare intensity map for the \ion{Fe}{12} $\lambda$195.12 line of EIS. We track the EUV bright points seen by both the instruments to obtain the offset. The image for the \ion{Ca}{2} H line of SOT is co-aligned with the pre-flare intensity map for the \ion{He}{2} $\lambda$256.32 line of EIS. We use the footpoints of the loops to obtain the offset. The line-of-sight (LOS) magnetic field measured by the Michelson Doppler Imager (MDI; \citealt{sche95}) aboard {\it SOHO} is co-aligned with the image for the \ion{Ca}{2} H line of SOT. We then use some small magnetic structures to obtain the offset. In addition, it is known that there exists an offset between the short waveband (SW) CCD, on which the \ion{Fe}{12} $\lambda$195.12 line is observed, and the long waveband (LW) CCD, on which the \ion{He}{2} $\lambda$256.32 line is observed. The offset value commonly used is 16\arcsec~in the N--S direction and 2\arcsec~in the W--E direction. Finally, data from all different instruments are co-aligned to that for the \ion{He}{2} line of EIS. Note that the difference in the viewing angles from different instruments can be safely ignored. In this paper, we use the coordinates of the EIS \ion{He}{2} line for all the images from other instruments.

\section{Data processing}\label{data}
\begin{figure}
\epsscale{1}
\plotone{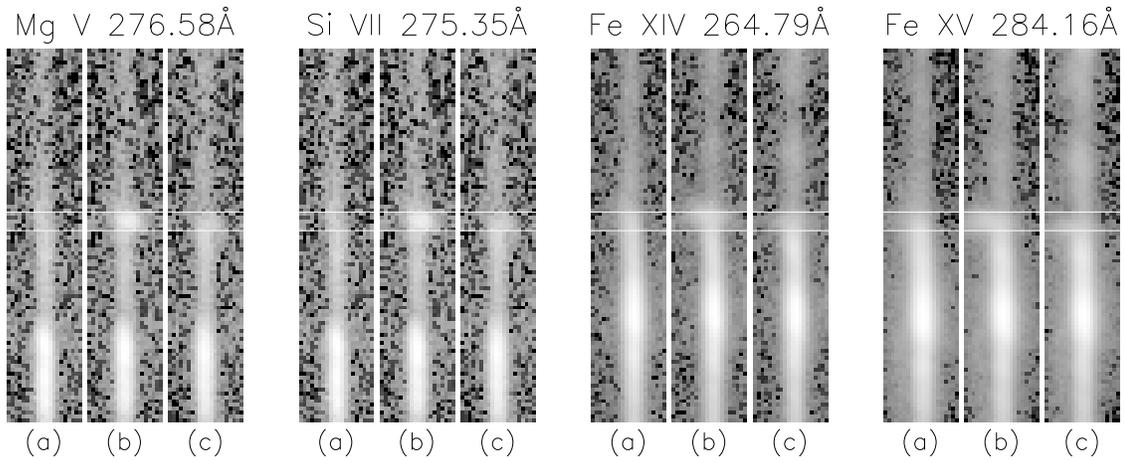}
\caption{Spectra observed by the EIS detector for four emission lines at 3 different times. The vertical axis is along the slit and the horizontal axis is for the wavelength dispersion. Subpanels (a), (b), and (c) represent spectra at 04:30:13, 04:35:35, and 04:38:16 UT, respectively. We bin the pixels in the slit direction between the two solid lines to increase the signal-to-noise ratio.}\label{fig2}
\end{figure}

Considering that the 2\arcsec~offset between the two CCDs of EIS in the W--E direction is non-negligible compared to the 5\arcsec~FOV in the scanning direction, we only select the emission lines from the LW CCD for this study, as listed in Table \ref{tab1}. Figure \ref{fig2} shows the spectrograms observed using EIS for four lines in time sequence. The vertical axis is along the slit direction and the horizontal axis is for the wavelength dispersion. There are intensity enhancements and red shifts in the \ion{Mg}{5} and \ion{Si}{7} lines, as well as co-spatial blue shifts in the \ion{Fe}{14} and \ion{Fe}{15} lines whose formation temperatures are much higher. To increase the signal to noise ratio, we bin the pixels in the slit direction between the two horizontal solid lines over which the line profiles do not change much. Furthermore, the binning is also done for every two columns in the scanning direction for the same purpose. The spectra from one binned pixel are then analyzed in detail. Figure \ref{fig3} shows the EIS \ion{He}{2} line intensity image, the EIT 195 \AA~image, the SOT \ion{Ca}{2} H image, and the LOS magnetic field measured by SOHO/MDI. The region between the vertical lines indicates the EIS FOV in the scanning direction. The cross in each panel indicates the location of the pixel to be studied in detail. Figure \ref{fig4} shows the spatial distribution of line intensity, Doppler velocity, and line width for the \ion{Fe}{12} $\lambda$195.12 raster before the flare.

First, a reliable rest wavelength measurement is critical to the measurement of the Doppler velocities of the evaporated plasma. It is known that the rest centroid of lines varies because of the EIS detector temperature variation during the {\it Hinode} orbit.  We set the rest wavelength using the quiescent region in the EIS raster from 01:15 to 03:27 UT and that from 03:27 to 04:19 UT. The average line center over the bottom 50\arcsec~in the slit direction is fitted by a sinusiodal function, which varies with time as
\begin{equation}\label{sine}
{\rm \varphi}=0.001996+0.01742\sin\left[\frac{\pi}{2942}(t-2132)\right],
\end{equation}
where $\varphi$ (in \AA) is the average line center subtracted by the default one, i.e., the line center listed in Table \ref{tab1}, and $t$ (in sec) is the time relative to 01:15:13 UT on 2007 December 7. The mean fitting error is 1.1 m\AA~, corresponding to $\sim$ 1.5 km s$^{-1}$ for the \ion{Fe}{12} $\lambda195.12$ line. The average line center and the fitting result for this line are shown in Figure \ref{fig5}. For lines not observed during 01:15--04:19 UT, the variation curve of the \ion{Fe}{12} $\lambda$195.12 line is applied to them, since it is the strongest and cleanest line in EIS observations (Harra, private communication).

Second, line blending is also important for the measurement of Doppler velocities, as well as the intensity and width of emission lines. For the lines that we select, the \ion{He}{2} $\lambda$256.32 line is blended with the \ion{Si}{10} $\lambda$256.37, \ion{Fe}{13} $\lambda$256.42, and \ion{Fe}{12} $\lambda$256.41 lines \citep{you07}. We find that the blending is evident in the region with coronal loops; however, the \ion{He}{2} line is the dominant component at loop footpoints and in quiescent regions. Therefore, the \ion{He}{2} line shows a profile that can be fitted by only one component in the region that we study.

Third, for most of the emission lines that we select, adopting a one-component Gaussian function can fit the line profiles very well. However, in some cases, there appears a blue-wing asymmetry in the \ion{Fe}{14} $\lambda$264.79 and \ion{Fe}{15} $\lambda$284.16 lines. We then need to apply a two-component Gaussian fit to these lines.

Finally, the function that we use to fit the line profile is
\begin{equation}\label{func}
\Psi=A+Bx+a_{0}\exp\left[-\frac{(x-a_{1})^{2}}{2a_{2}^{2}}\right],
\end{equation}
where $x$ (in \AA) is the wavelength, $a_{0}$ (in ergs cm$^{-2}$~s$^{-1}$~sr$^{-1}$~\AA$^{-1}$) is the peak value of the profile, $a_{1}$ (in \AA) is the line center, $a_{2}$ (in \AA) is the line width, and A and B are constants for the linear background. When we perform a multi-component fit, the function becomes
\begin{equation}\label{multi}
\Psi=A+Bx+\sum_{i=1}^{n} a_{i,0}{\rm exp}\left[-\frac{(x-a_{i,1})^{2}}{2a_{i,2}^{2}}\right],
\end{equation}
where $n$ is the number of Gaussian components required.

We perform fits to 10 emission lines for 9 rasters covering the rise and decay phases of the flare. The details of the physical parameters deduced from the fits are presented in the following sections.

\section{Results}\label{result}
\subsection{Temperature-dependent evaporation velocity}\label{velocity}
Figure \ref{fig6} shows the line profiles from the binned pixels at $\sim$~04:35 UT around the point marked by the cross in Figures \ref{fig3} and \ref{fig4}. It is clear that all the lines, except for the \ion{Fe}{15} $\lambda$284.16 line, show a symmetric Gaussian profile. For the case of the \ion{Fe}{15} line, we need to use a two-component fit to properly account for the blue-shifted component.\footnote{In the two-component fit, the stationary component refers to the component that is less Doppler-shifted. In most cases, this component has a velocity of 10--20 km s$^{-1}$. Note that the rest wavelength of the line can be determined through the sinusoidal fit as described in \S \ref{data}.} Several emission lines may exist in the wavelength window of a line, e.g., the \ion{Fe}{10} line. In this case, a multi-component Gaussian fit (one component for each line) is applied for the wavelength window.

From Figure \ref{fig6}, we find that the \ion{Si}{7} line and those formed at lower temperatures show red shifts; while the \ion{Fe}{10} line and those formed at higher temperatures show blue shifts. The blue shift in the \ion{Fe}{15} $\lambda$284.16 line yields an upward velocity of $\sim$~100 km s$^{-1}$. The velocities for the 10 lines are plotted as a function of their formation temperature in Figure \ref{fig7}. Note that the \ion{Si}{7} and \ion{Mg}{7} lines are formed at nearly the same temperature; therefore, their Doppler velocities are very close, as expected. A similar behavior is found for the \ion{Si}{10} and \ion{S}{10} lines. A cubic polynomial and a linear function are used to fit the upward and downward flow velocities, respectively, as a function of their formation temperature. The results show that the lines from \ion{He}{2} to \ion{Si}{7} are red shifted by $v_{down}=33.4-22.7~T$ and those from  \ion{Fe}{10} to \ion{Fe}{15} are blue shifted by $v_{up}=246-538~T+370~T^2-83.9~T^3$, where the quantities {\it v} and {\it T} are in units of km~s$^{-1}$~and MK, respectively. The fitting results are shown as dashed lines in Figure \ref{fig7}. Note that adopting a polynomial function of higher orders can fit all the velocity points, both positive and negative. As a test, we find a quartic polynomial, $v=28.9+35.8~T-143~T^2+106~T^3-25.1~T^4$, can fit all the points well except for that of \ion{Fe}{10}. However, doing so only makes sense mathematically without increasing the physical significance.

\subsection{Temporal evolution of the evaporation}\label{evolution}
Fortunately, the EIS observations for this study cover the initial, the impulsive, and the decay phases of the flare. We thus  investigate the temporal variation of the line intensities, Doppler velocities, and line widths for 9 emission lines from \ion{He}{2} to \ion{Fe}{15}. Since there are many lines in the window of the \ion{Fe}{10} $\lambda$257.26 line, it is difficult to measure and fit this line accurately in some cases. Therefore, the \ion{Fe}{10} $\lambda257.26$ line is not selected for the study of temporal evolution behaviors. However, this line is still included in the study of temperature-dependent velocities, since it is the only line formed at around 1.0 MK. Note that the \ion{He}{2} line is blended with some other lines as discussed in \S \ref{data}. However, we find that the line shows a rather symmetric Gaussian profile, implying the dominance of the main component, especially at the loop footpoints. Thus the \ion{He}{2} line is included in the following investigation.

Figure \ref{fig8} shows the temporal variation of the line intensities, Doppler velocities, and line widths for the 9 emission lines. Note that for the \ion{Fe}{15} line, the parameters for both the blue-shifted component and the stationary component are plotted. The intensities of all the emission lines increase significantly in the impulsive phase. This is generally consistent with the scenario predicted by the evaporation model \citep{fish85a,fish85b,fish85c,liu09}, in which the dense chromospheric plasma is heated to temperatures similar to those in the transition region and the corona, resulting in intensity enhancements of emission lines formed at these temperatures. For the \ion{Fe}{15} line, in particular, its two-component feature seems to contradict theoretical model that predicts no stationary component during the impulsive phase of a flare. However, the emission of the \ion{Fe}{15} line is dominated by the enhanced blue-shifted component, while the emission from the stationary component of the \ion{Fe}{15} line only slightly increases during the impulsive phase. Therefore, the blue-shifted component is clearly flare-related, while the stationary component observed in the impulsive phase seems to be mostly from a non-flare emission source.

In the decay phase of the flare, the intensity of each line generally decreases but remains above their pre-flare value for almost an hour. The plots of \ion{He}{2}, \ion{Si}{10}, \ion{S}{10}, \ion{Fe}{14}, and \ion{Fe}{15} line intensities as a function of time reveal later second peaks during $\sim$~04:40--04:50 UT (see Figure \ref{fig8}, the first and the last 4 panels in the left column). This later peak in the decay phase could be induced by hotter plasma cooling back down into coronal passband. The similarity between the \ion{He}{2} and the coronal lines may be attributed to the coronal lines blended with the \ion{He}{2} line, as discussed in \S \ref{data}.

In particular, the evolution of the \ion{Fe}{15} line is more interesting. The later peak in this line is mainly from the stationary component and much higher than the pre-flare intensity; the blue-shifted component contributes $\sim 23\%$ of the total intensity. Note that the stationary component is not absolutely static, as mentioned above. It has a velocity of about 10--20 km s$^{-1}$. The plasma being cooled down near the top of coronal loops, which is probably overlapped with the footpoint along the line of sight, could contribute to the enhancement of this low-velocity component. A more accurate determination of the source regions of the blue-shifted and stationary components is limited by the small FOV of EIS.

Recently, \citet{pet10} found that the Fe XV line always exhibits a blue-wing asymmetry in active regions. Therefore, the blue-shifted component may be partly contributed by flows that are not related with the flare. We should then be cautious about explaining the fitting results especially in the later stage of the flare. However, in the impulsive phase, the intensity of the blue-shifted component is significantly enhanced. The fitting results can reasonably reflect the plasma dynamics that is caused by the flare itself.

The widths of most of the lines follow a pattern that is well consistent with that for line intensities, i.e., a significant increase in the impulsive phase and a gradual decrease in the decay phase. One may notice that this pattern is not evident for the widths of the \ion{Mg}{5}, \ion{Mg}{6}, and \ion{Mg}{7} lines. Note that the signal to noise ratio in these lines is lower than that in other lines, except for the time period in and shortly after the impulsive phase, when the line intensity is the most enhanced. Therefore, the measurement for these lines may be less reliable than that for other lines, as revealed by the larger error bars at some data points for the \ion{Mg}{6} line. Nevertheless, these uncertainties do not conceal the general scenario, i.e., the significant line broadening in the impulsive phase.

\section{Discussions and Summary}\label{discu}
We report a case of an explosive chromospheric evaporation during the impulsive phase of a microflare. We calculate in detail of the line intensity, the Doppler velocity, and the line width for 10 emission lines formed from 0.05 to 2.5 MK. For 9 out of the 10 lines, we study further the temporal evolution. Our key findings are summarized as follows.
\begin{enumerate}
\item{We measure Doppler velocities over an area in the flare ribbon. The emission lines formed in the temperature range of 0.05--0.63 MK are red shifted and those in the range of 1.0--2.5 MK are blue shifted, implying a clear case of explosive chromospheric evaporation.}
\item{The upflows and the downflows show a dependence on temperature, i.e., $v_{up}=246-538~T+370~T^2-83.9~T^3$ and $v_{down}=33.4-22.7~T$, where the quantities {\it v} and {\it T} are in units of km~s$^{-1}$~and MK, respectively. The transition between upflows and downflows lies in a very narrow temperature range of $< 0.3$ MK.}
\item{Both a significant line intensity enhancement and a line broadening are found and correlated with the development of the flare. The \ion{Fe}{15} $\lambda$284.16 line, the hottest line in this observation, show a blue-wing asymmetry in the impulsive phase. Thus, the line profile comprises of a relatively stationary component and a blue-shifted component. The latter disappears in the late decay phase.}
\end{enumerate}

\begin{figure}
\epsscale{1}
\plotone{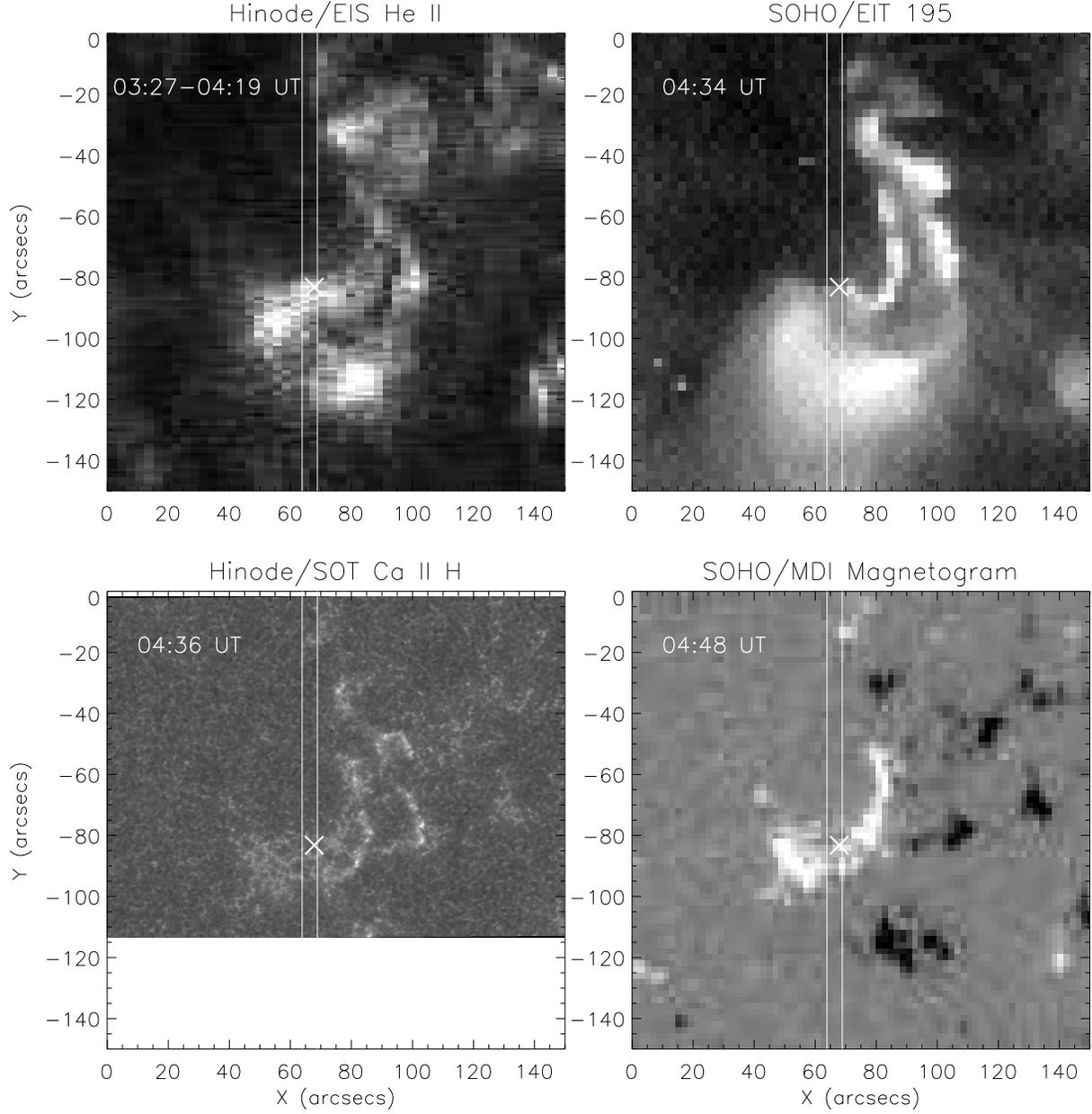}
\caption{ EIS intensity map for the \ion{He}{2} $\lambda256.32$ line observed from 03:27 to 04:19 UT prior to the flare occurrence, {\it SOHO}/EIT 195 \AA~image, {\it Hinode}/SOT \ion{Ca}{2} H image, and magnetogram from {\it SOHO}/MDI. The flare onset was at around 04:30 UT. All the images are co-aligned to the EIS intensity map. The vertical lines indicate the EIS FOV in the scanning direction. The cross in each panel indicates the spatial pixel that is studied in detail.}\label{fig3}
\end{figure}

\begin{figure}
\epsscale{1}
\plotone{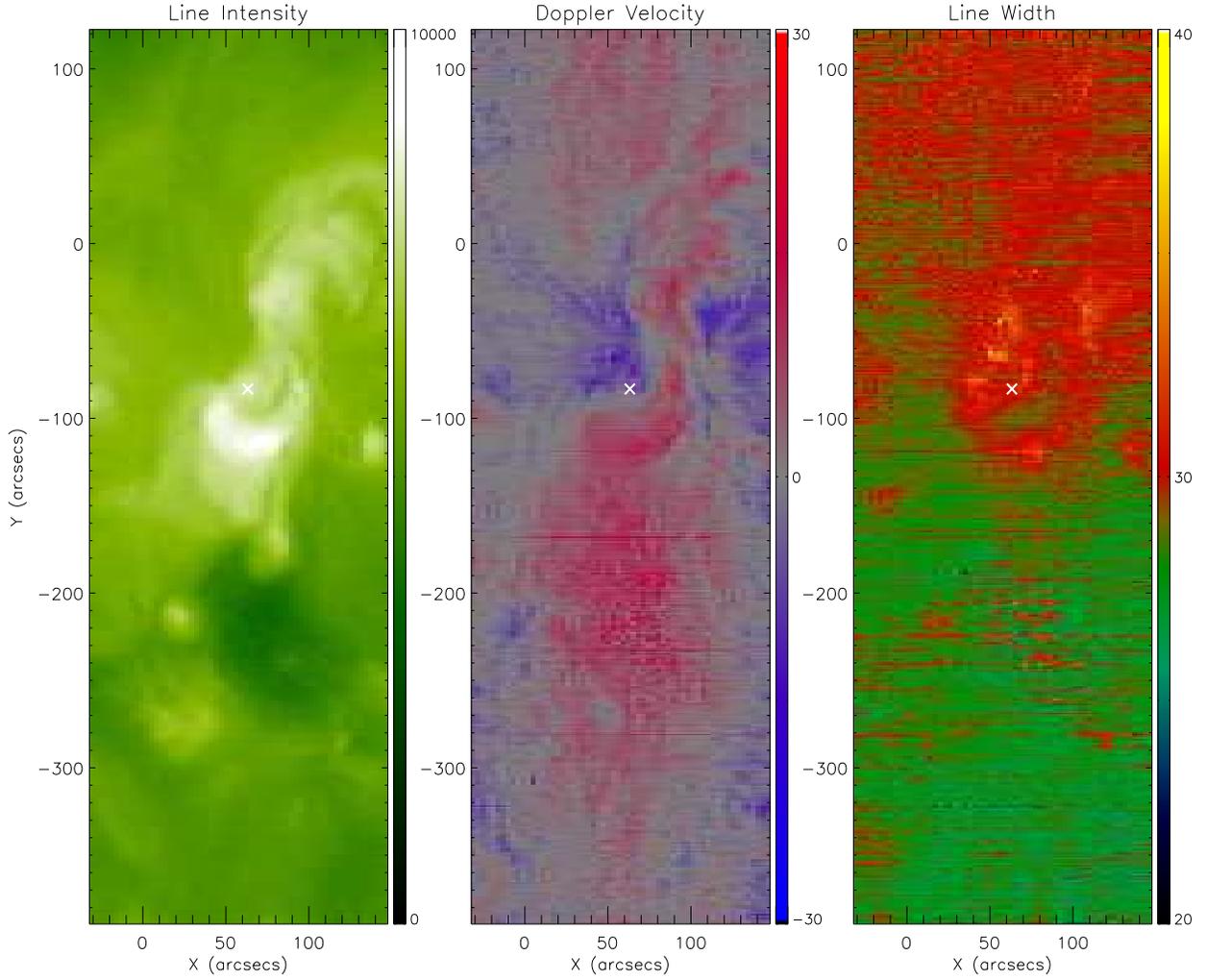}
\caption{Line intensity, Doppler velocity, and line width derived from the \ion{Fe}{12} $\lambda$195.12 line for the active region. The time range for scanning is from 03:27 to 04:19 UT, which is before the flare occurrence. The cross in each panel indicates the spatial pixel that is studied in detail.}\label{fig4}
\end{figure}

\begin{figure}
\epsscale{1}
\plotone{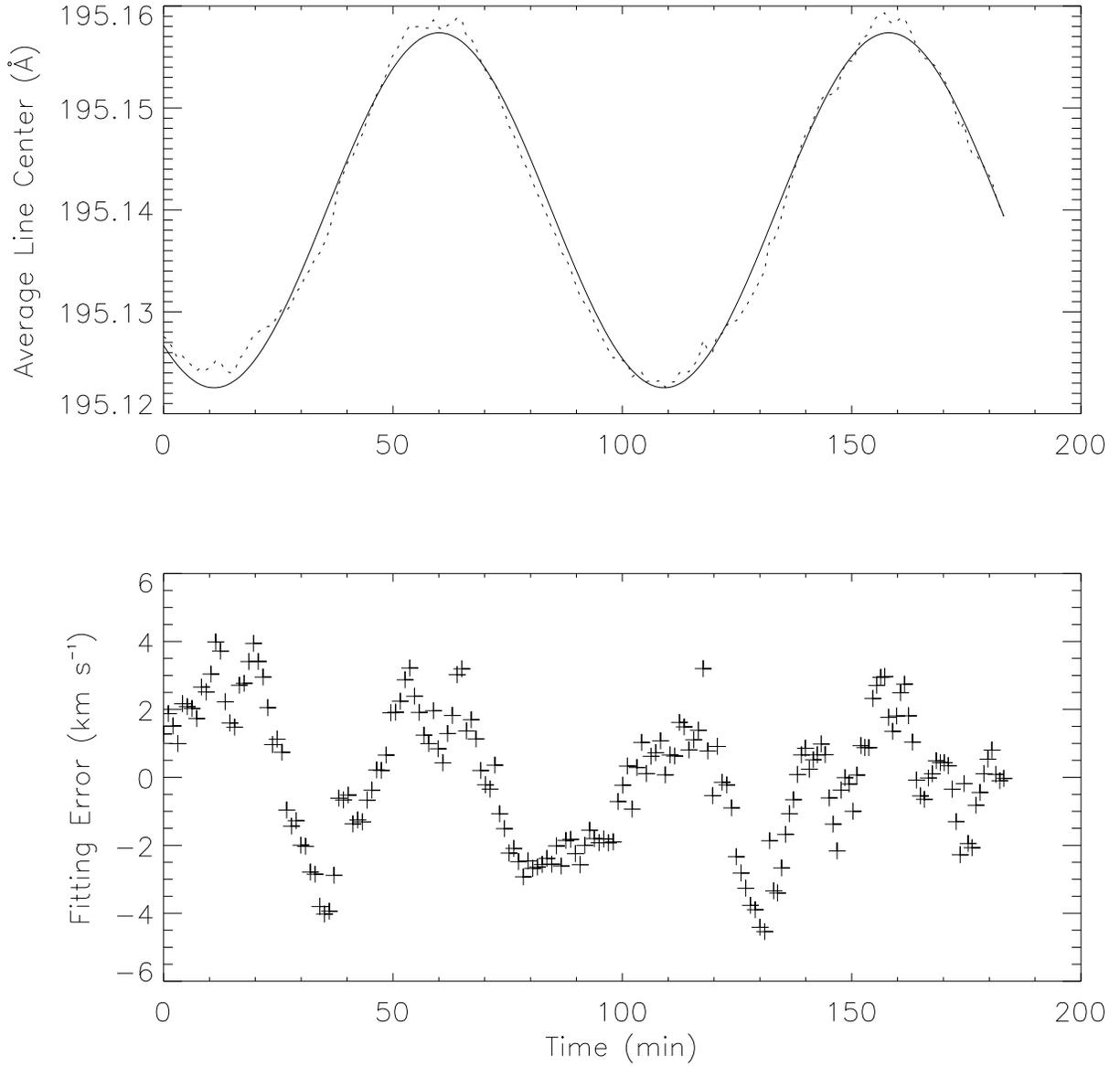}
\caption{Correction curve for the line center variation of the EIS emission lines. Top panel: the average line center measured from the quiescent regions for the \ion{Fe}{12} $\lambda$195.12 line (dotted line) and the sinusoidal fit for it (solid line). Bottom panel: the difference between the two curves shown in the top panel. The time is relative to 01:15 UT. }\label{fig5}
\end{figure}

\begin{figure}
\epsscale{1}
\plotone{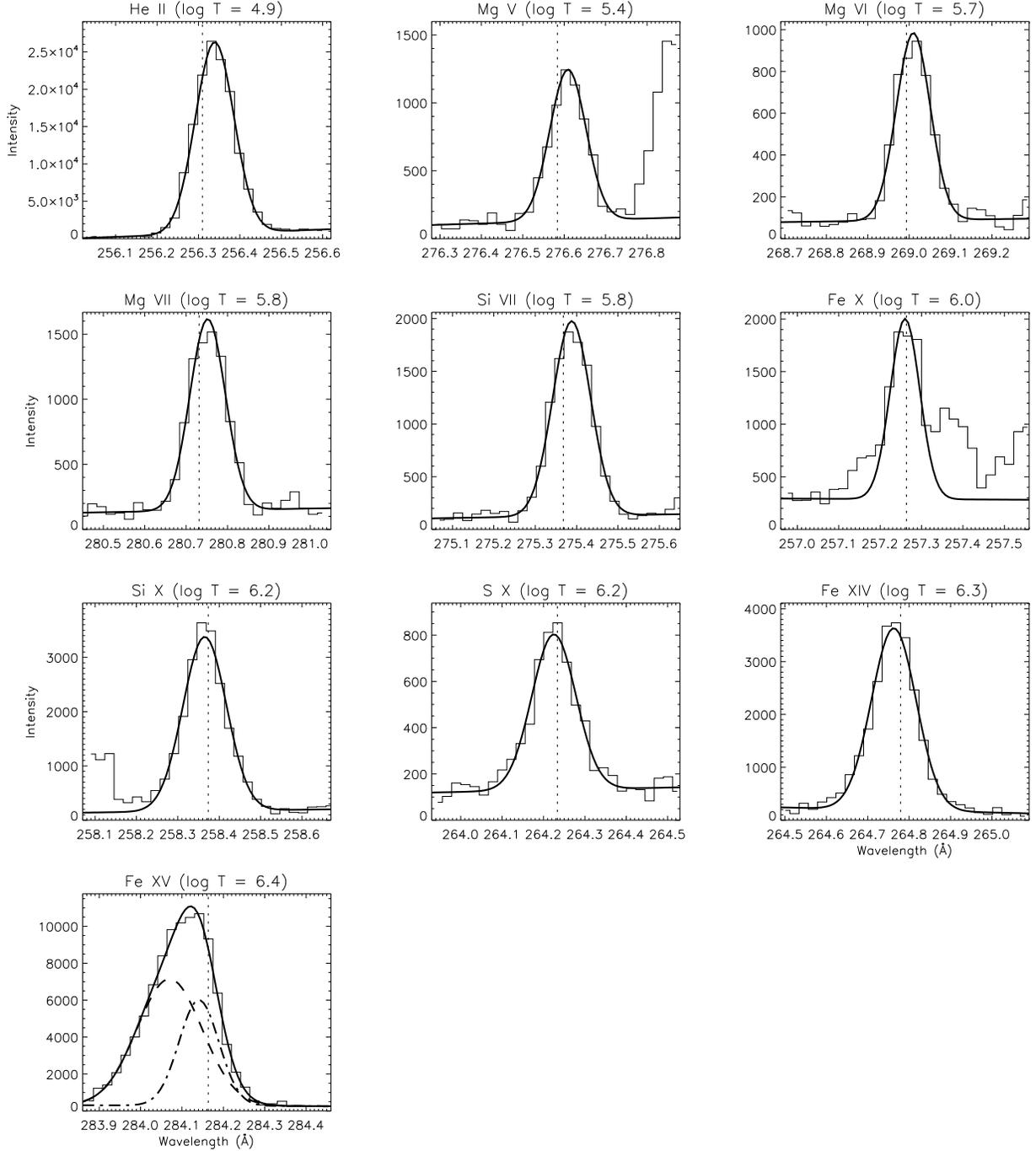}
\caption{Line profiles and fitting results for 10 emission lines for the spatial pixel (shown in Figures \ref{fig3} and \ref{fig4}) at $\sim$~04:35 UT. The intensity is in units of ergs cm$^{-2}$~s$^{-1}$~sr$^{-1}$~\AA$^{-1}$. In each panel, the histogram is for the observed line profile and the vertical dotted line is for the rest wavelength calculated from the sinusoid shown in Figure \ref{fig5}. The first 9 panels show the one-component Gaussian fits by the solid lines. The last panel shows the two-component Gaussian fit, where the dashed line is the blue-shifted component, the dot-dashed line is the stationary component, and the solid line is the sum of them.}\label{fig6}
\end{figure}

The most meaningful result is that we find explosive chromospheric evaporation in a microflare, with a division between upflows and downflows at a temperature of $\sim$ 0.7--1.0 MK. Explosive evaporation has previously been observed in C-class and larger flares and was found to be cospatial and cotemporal with HXR emissions \citep{mil06a,mil09}. In the event presented here, the upflows shown in hotter lines are definitely cospatial and cotemporal with the downflows shown in cooler lines, consistent with previous studies. Furthermore, we find that the mass flow velocities are temporally correlated with the derivative {\it GOES} 1--8 \AA~light curve, i.e., the proxy of HXR light curve. The intensities of these lines are significantly enhanced in the impulsive phase and quickly decrease after the flare peak time. Therefore, the evaporation is likely induced by a nonthermal electron beam during the impulsive phase, as predicted by the theoretical electron beam heated model \citep{fish85a, fish85b, fish85c}. \citet{fish85a} found an energy flux of $10^{10}$~ergs~cm$^{-2}$~s$^{-1}$ as a threshold for the explosive evaporation, which was later verified by observations of \citet{mil06a,mil06b} and \citet{mil09}. Our observations imply that the energy flux of the electron beam in a microflare may be comparable to that in much bigger events, though the beam should be restricted to a rather smaller area, resulting in a very small integrated flux as observed in this B-class flare. It is worth mentioning that \citet{hann08} found nonthermal emission in remarkably high energies ($>$ 50 keV) in an A-class microflare, showing direct evidence of nonthermal electron beam heating in a microflare. We notice that the velocities of the downflows (20--30 km s$^{-1}$) and the upflows (100 km s$^{-1}$) are somewhat smaller than typical values in more energetic events. This may partly be due to the insufficient spatial resolution of the EIS scanning observations, in which the observed line profiles are in deed a convolution of that in the heated region and that in the nearby quiscent region. Owing to the smaller EIS FOV in this study, we are unable to estimate the exact spatial scale of the evaporation, except that it is $\sim$~10\arcsec~in the N--S direction. We suspect that the brighter flare ribbon to the west of the EIS FOV (see Figure \ref{fig3}, left panel) may exhibit even more significant evaporation than what we observe in the EIS FOV. By comparison, \citet{mil08} studied a microflare using EIS and {\it RHESSI} observations. They found only a weak blue shift of 14 km s$^{-1}$ in the \ion{Fe}{15} line in the flare ribbon without any cospatial HXR emission. Moreover, \citet{broh09} found a gentle chromospheric evaporation in the impulsive phase of a microflare, which has also no detectable HXR emission. Therefore, our event implies the possible existence of a small-scale strong electron beam in microflares. For comparison, \citet{jess08} detected a strong white-light emission, though restricted to a very small area, in a relatively small flare. Note that the white-light emission, like the explosive evaporation scenario, is also thought to be powered by a strong electron beam in the conventional point of view \citep{hud72}.

\begin{figure}
\epsscale{1}
\plotone{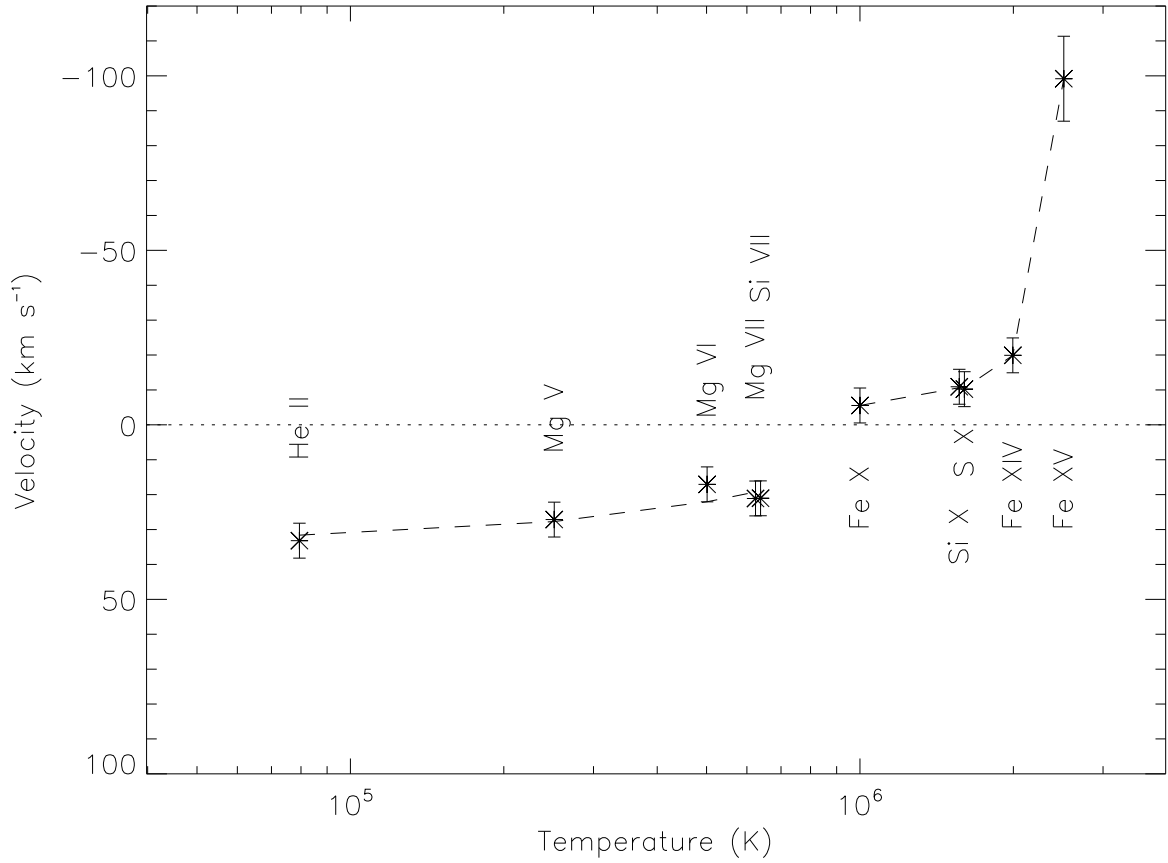}
\caption{Velocities of evaporated plasma as a function of line formation temperature for the selected spatial pixel at $\sim$~04:35 UT. The dashed lines are a linear fit to velocities from lines of 0.05--0.63 MK and a polynomial fit to velocities from lines of 1.0--2.5 MK, respectively.}\label{fig7}
\end{figure}

\begin{figure}
\epsscale{0.8}
\plotone{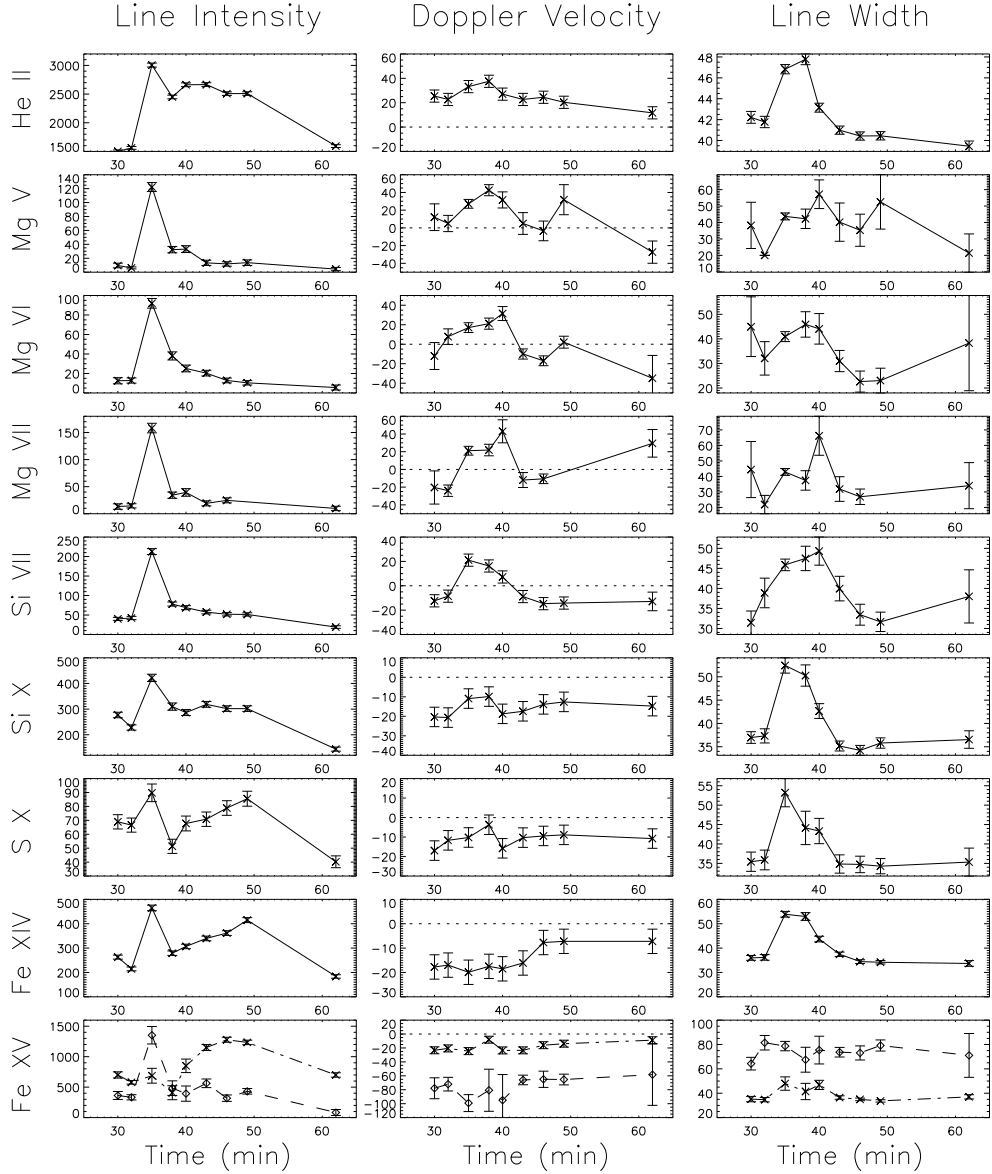}
\caption{Time variations of the line intensities integrated over wavelength (in ergs cm$^{-2}$~s$^{-1}$~sr$^{-1}$), Doppler velocities (in km s$^{-1}$), and line widths (in m\AA) for the 9 emission lines. In the first 8 rows, we fit the line profiles with a single Gaussian component only. In the last row, we use the two-component fit, where the dashed line is for the blue-shifted component and the dot-dashed line is for the stationary component. The time is relative to 04:00 UT.}\label{fig8}
\end{figure}

\citet{ima07} reported temperature-dependent velocities in the decay phase of an X-class flare. They found that the upward velocities in lines of below 1.0 MK show a weaker dependence on temperature, and those in lines of higher than 1.0 MK show a stronger dependence on temperature. In our event, however, the cooler lines show downward velocities. In addition, we observe a significant temperature dependence of upflows (downflows) during the impulsive phase rather than in the decay phase. Thus, the situations in these two events are quite different. Our observations are more like that of \citet{mil09}, who studied a C-class flare and found temperature-dependent velocities in the impulsive phase. However, the transition temperature between upflows and downflows in our event is $\sim$~0.7--1.0 MK which is lower than the value obtained by \citet{mil09}, i.e., $\sim$~1.5--2.0 MK. Theoretically, the existence of temperature-dependent velocities was predicted by \citet{fish85c}, who considered a single burst of energetic electrons, and by \citet{liu09}, who adopted a continuous electron deposition in flare dynamic models. We find that the distribution of velocities, as shown in Figure \ref{fig7}, is quite consistent with the simulation results of \citet{liu09} (see their Figure 10, panel c). Thus, the upward and downward velocities may be interpreted as follows. Energetic electrons deposit energy in the chromosphere, where the local pressure is enhanced to exceed that of the overlying corona. Such a pressure gradient can drive the heated plasma both upward and downward. The short-lived EUV brightenings, which last for about 5 minutes, are due to either a short duration of electron injection or a high radiative cooling rate in the condensed plasma.

We study in detail the temporal evolution of the line intensities and widths for the 9 emission lines, as well as their Doppler velocities, using the high spectral resolution EIS observations. The line intensities are significantly enhanced in the impulsive phase of the flare, accompanied by the temperature-dependent velocities as mentioned above. We find a later second peak in the decay phase ($\sim$~04:40--04:50 UT) for intensities of coronal lines, in contrast to the intensity decrease observed in the cooler lines. The most likely explanation could be the hotter plasma with temperatures much higher than 2 MK cooling back down into the coronal passband. Unfortunately, the \ion{Fe}{15} line is the hottest line in this observation, thus we are unable to implement a direct comparison between the light curves of coronal lines and those of lines formed in much higher temperatures. Note that such an intensity increase in the decay phase can now be cheched by the Extreme ultraviolet Variability Experiment (EVE; \citealt{woo10}) instrument on {\it Solar Dynamics Observatory} ({\it SDO}).

We also  find a significant line broadening, i.e., the line width increases by $\sim$ 60\%, in association with the intensity enhancement. By checking carefully the line profiles pixel by pixel, we confirm that the line broadening during the impulsive phase is of physical significance. There are two interpretations of the line broadening. On one hand, they can be interpreted as an enhancement of the nonthermal velocity induced by certain physical processes. For example, \citet{chen10} reported that the line widths increase by 12\% at the outer edge of the EIT wave-associated dimming; \citet{mcin09} interpreted the line broadenings in the CME-induced dimming as the growth of Alfv\'en wave amplitudes. On the other hand, an insufficient resolution is also a possible cause. If there exist spatially unresolved components in each pixel that are heated differently and exhibit different mass flows, the spatially integrated line profile should be more broadened than a single one in each component. For instance, \citet{dosc08} proposed an origin of the line broadening due to the multiplicity of flows. We think that either one or both could work, i.e., the velocity dispersion along the line of sight and/or the spatial difference lead to the observed line broadening.

In summary, we report clear evidence of explosive chromospheric evaporation in a B-class microflare. Temperature-dependent upward and downward velocities and line broadenings are observed in the impulsive phase of the flare. This event  implies that the evaporated flows could possibly be induced by energetic nonthermal electrons. The evidence, however, is not very solid, since there are no observations from {\it RHESSI} for this event. Nevertheless, our findings provide a scenario in which a small flare may be locally ``very energetic" and possesses physical processes that were ever thought to exist only in bigger flares.

\acknowledgments
 We thank the referee for constructive comments that helped to improve this manuscript. This work was supported by NSFC under grants 10828306 and 10933003 and by NKBRSF under grant 2006CB806302. The authors thank Ryan Milligan and Louise Harra for insightful discussions, comments and helpful suggestions. Hinode is a Japanese mission developed and launched by ISAS/JAXA, collaborating with NAOJ as a domestic partner, NASA and STFC (UK) as international partners. Scientific operation of the Hinode mission is conducted by the Hinode science team organized at ISAS/JAXA. This team mainly consists of scientists from institutes in the partner countries. Support for the post-launch operation is provided by JAXA and NAOJ (Japan), STFC (U.K.), NASA, ESA, and NSC (Norway).

\clearpage

\clearpage

\clearpage

\clearpage

\clearpage

\clearpage

\clearpage

\clearpage

\clearpage

\begin{deluxetable}{lcc}
\tablecolumns{3}
\tablewidth{0pc}
\tablecaption{EIS lines used in this study}
\tablehead{
\colhead{Ion} & \colhead{~~Wavelength (\AA)} & \colhead{Log $T_{max}$ (K)}
}
\startdata
\ion{He}{2}&256.32&4.9\\
\ion{Mg}{5}&276.58&5.4\\
\ion{Mg}{6}&268.99&5.7\\
\ion{Mg}{7}&280.75&5.8\\
\ion{Si}{7}&275.35&5.8\\
\ion{Fe}{10}&257.26&6.0\\
\ion{Si}{10}&258.37&6.2\\
\ion{S}{10}&264.23&6.2\\
\ion{Fe}{14}&264.79&6.3\\
\ion{Fe}{15}&284.16&6.4\\
\enddata\label{tab1}
\end{deluxetable}

\end{document}